**Generalized asset pricing:**
**Expected Downside Risk-Based Equilibrium Modelling**

Mihály Ormos [a], Dusán Timotity [a]

[a]Department of Finance, Budapest University of Technology and Economics

This version: October 20, 2015

**Abstract**

We introduce an equilibrium asset pricing model, which we build on the relationship between a novel risk measure, the Expected Downside Risk (*EDR*) and the expected return. On the one hand, our proposed risk measure uses a nonparametric approach that allows us to get rid of any assumption on the distribution of returns. On the other hand, our asset pricing model is based on loss-averse investors of Prospect Theory, through which we implement the risk-seeking behaviour of investors in a dynamic setting. By including *EDR* in our proposed model unrealistic assumptions of commonly used equilibrium models – such as the exclusion of risk-seeking or price-maker investors and the assumption of unlimited leverage opportunity for a unique interest rate – can be omitted. Therefore, we argue that based on more realistic assumptions our model is able to describe equilibrium expected returns with higher accuracy, which we support by empirical evidence as well.

**Acknowledgements:** We are grateful to the seminar and conference participants at the Workshop on Behavioural Economics and Industrial Organization at Corvinus University, 2014, 10th EBES Conference in Istanbul, the 10th International Scientific Conference on European Financial Systems at Masaryk University and 5th International Conference "Economic Challenges in Enlarged Europe" in Tallinn.
We would like to gratefully acknowledge the valuable comments and suggestions of the Editor, Prof. Sushanta Mallick and three anonymous referees that contribute to a substantially improved paper.
Mihály Ormos acknowledges the support by the János Bolyai Research Scholarship of the Hugarian Academy of Sciences. Dusán Timotity acknowledges the support by the Fundation of Pallas Athéné Domus Scientiae. This research was partially supported by Pallas Athene Domus Scientiae Foundation.

## 1. INTRODUCTION

The most commonly applied asset pricing model, the Capital Asset Pricing Model (CAPM) (Lintner, 1965; Mossin, 1966; Sharpe, 1964), defines three sets or areas of constraints or boundary conditions: (1) there is a perfect market, which consists of four assumptions: (a) investors have constant preferences, (b) are price-takers, (c) are perfectly informed and (d) there is no transaction cost; (2) conditions of investors' behaviour: (a) they are rationally risk-averse, therefore they hold efficient portfolios defined by Markowitz and (b) due to their rationality and perfect information have homogenous expectations; (3) assumptions of investment possibilities: (a) in addition to risky portfolios, investors may invest in risk-free assets as well and (b) they can borrow infinite amounts at this risk-free rate. Furthermore, the standard regressions behind the theoretical model assume that the returns are normally distributed. Using our proposed equilibrium asset pricing model that is based on loss-averse investors described by Prospect Theory (Kahneman and Tversky, 1979) rather than Expected Utility Theory (EUT) (Von Neuman and Morgenstern, 2007) the assumptions of risk-averse and price-taker investors, normal distribution of returns (thus the required linear regression between risk and expected return (Erdős et al., 2011) and unlimited borrowing at a unique risk-free interest rate can be omitted.

In our model based on Kahneman and Tversky's (1979) Prospect Theory, we apply the assumption that in some cases investors define a reference point different from zero on their utility curve, resulting in anchoring (Ariely et al., 2003). We show that in some cases they do not refuse risk, and moreover they start to follow risk-seeking behaviour to an extent, since their expected utility can be maximized with this behaviour; thus, the assumption of risk-averse investors cannot be valid. However, we show that this type of behaviour can be implemented in standard asset pricing without any problems, which yields that the inclusion of risk-seeking investors does not change the conclusion of equilibrium models. We approximate the perceived risk by a novel measure, the Expected Downside Risk (*EDR*), which is based on Value-at-Risk (Campbell et al., 2001; Jorion, 2007) and Conditional Value-at-Risk (Rockafellar and Uryasev, 2000). Through this method we can define the expected loss below the expected return weighted by its probability. The results we present in this paper are significantly different from those of well-known models that approximate expected return by standard deviation, such as Modern Portfolio Theory (Markowitz, 1959). Although the mean-variance optimization is valid under ellipctical distributions as well (Chamberlain, 1983) and the normality condition has already been omitted in some regressions (such as the Markowitz 2.0 model based on Conditional Value-at-Risk (Kaplan, 2012) or Iglesias (2015)), these models still require unrealistic assumptions and miss a coherent and general approach to modern asset pricing, which we aim to introduce with our approach. In the followings we implement the risk-seeking behaviour in an *EDR expected return* environment by combining the risk-seeking behaviour and our proposed risk meausre.

Furthermore, we show the effect of limited borrowings, where different leverage constraints and interest rates are applied for every single investor. Hence, a more realistic and precise way to explain the individual optimization method is defined, which yields the unnecessity of the assumption of unlimited leverage opportunity for a unique risk-free rate.

Subsequent to describing the individual optimization the aggregation method is discussed (i.e. how the market sets the expected return of a given asset). Here we include price-maker investors as well, who may have a significant effect on price formation by block transactions or in illiquid market segments.

Finally, we provide a pricing equation that approximates the expected return by *EDR* and compare it against alternative risk measures.

The paper is structured as the following: section 2 discusses risk-seeking behaviour and its implications in asset pricing models, then our proposed risk measure is defined in section 3. In section 4 these two are combined together and risk-seeking behaviour is shown in an *EDR* setting. Section 5 is related to restricted borrowing limits and their effect on portfolio choice. Section 6 implements all the previous findings in one model and describes the formulation of the expected return including the pricing equation. Finally, the paper ends with a brief conclusion. All sections are divided into theoretical and empirical subsections where empirical evidence is provided to facilitate understanding the main ideas and to support the theory.

## 2. THE CAUSE OF RISK-SEEKING BEHAVIOUR

### 2.1 Risk aversion or loss aversion

According to Expected Utility Theory (EUT) (Von Neuman and Morgenstern, 2007), perceived utility is a concave function of total wealth. This leads to the law of diminishing marginal utility, which means that the marginal utility that a person derives from consuming each additional unit of a product declines. According to the theory, this utility perception causes risk aversion, meaning that a mathematically fair investment with equal expected gain and loss would have a negative effect on expected utility.

There are two main risk aversion factors applied in economic theory: constant absolute risk aversion (CARA) and constant relative risk aversion (CRRA) (Pratt, 1964). The first one describes the risk premium necessary to invest in a mathematically fair investment, which does not depend on the reference wealth. The second one states that risk aversion changes with the change of wealth in a constant way, meaning that reference wealth multiplied by CARA is constant over time for each investor. Both measures have advantages and disadvantages over the other; however, in the case of small changes of wealth, we can use both of them for any investor. According to Markowitz (1959), either examining the utility curve for CARA or for CRRA, the approximation of the utility of investment *F(U(F))* using a Taylor series is defined as:

$$U(F) \cong E(F) - 0.5a\sigma^2 \qquad (1)$$

where *U(F)* is the expected utility of an investment *F*, *E(F)* its expected value, *α* the Arrow-Pratt measure of constant absolute risk aversion and $\sigma^2$ the variance that measures the risk of investment *F*. This is a key step in our model since there is a debate going on about which utility function to use in Prospect Theory. Although Kahneman and Tversky (1979) propose applying a power function similar to the one used in the CRRA approach, numerous studies show that an exponential function performs better (Smidts, 1997) or at least equally well (Beetsma and Schotman, 2001). Since CARA is independent of wealth, it makes it easier to describe the risk attitude of a given investor; therefore, we apply its corresponding exponential utility function in our model.

By accepting the assumptions that $\sigma^2$-*E(r)* variance-expected return efficient combinations can be described with a concave curve with positive slope – or in the case of unlimited borrowings with constant slope – and that investors are risk-averse (hence their CARA is positive, thus their utility curve is convex and monotonically increasing), optimal choice can be defined for any investor.

However, according to the Prospect Theory of Kahneman and Tversky (1979), investors do not behave this way, and their actions cannot be explained by Expected Utility Theory (EUT). The authors underline that investors' decision-making process is based not solely on economic rationality but subjective elements as well, which motivates them to behave in a way that would be completely irrational in standard economic theories. According to Shefrin (2002), these behavioural patterns are heuristic-driven biases and frame dependencies. There are numerous heuristics that play important roles in decision-making, such as representativeness (that is, people overestimate the frequency of events surrounding them and they apply stereotypes in their decisions), availability bias (that is, people rely too heavily on their own experience, and easily available information) or – most importantly in this paper – anchoring (that is, people make reference points and adjust new information according to them). These subjective elements cause biases in investors' expected probabilities, therefore they overestimate (and overreact to) rare events and underestimate (underreact to) frequent events.

Based on the heuristics mentioned above, Kahneman and Tversky develop a model that – in spite of EUT, which defines utility as the function of total wealth (the absolute way) – measures utility through the change of wealth (in a relative way), which prove to be a more precise method to describe investors behaviour. According to their theory, positive changes of wealth can be described with a function similar to the EUT utility function (which is concave and has monotonic growth), hence the law of diminishing marginal utility (Gossen, 1854) stays intact. However, in the case of negative changes of wealth (losses), although investors keep being risk-averse in normal cases (since their reference point returns to zero and the slope of the curve on the loss side is on average 2.25 times the slope on the gain side, thus their utility decreases 2.25 times more for x loss than it increases for x gain), this utility function becomes convex. Therefore, they define an S-shaped value function (utility curve) (Figure 1). This value function, however, could make current asset pricing models useless by introducing the anchoring heuristic (i.e. including previous outcomes in utility perception) as risk-seeking behaviour emerges this way and optimization with positive CARA is no longer available (i.e. no optimal portfolio with finite risk is present).

**Figure 1: Utility curve in Prospect Theory**

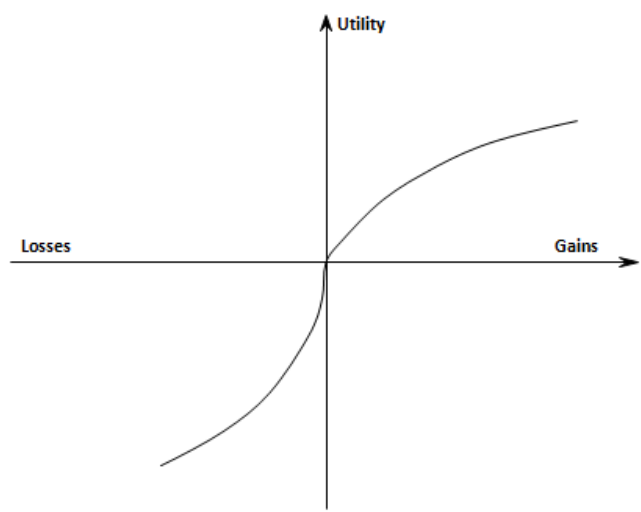

**Source:** Kahneman and Tversky (1979, p.279)

## 2.2 The emergence of risk-seeking

The Kahneman and Tversky (1979) model enables analysis of single period choices. According to their theory, the investor makes its choice from the zero reference point every time; however, in reality this is not always true. As Shefrin (2002) argues, the heuristic-driven biases, for different reasons – such as the anchoring effect (adjusting to a reference situation), the inclusion of sunk costs or the disposition effect (where investors give up properly managing their portfolios after a massive loss and do not sell) – may motivate investors to fix a reference point other than zero on their value function and hold that at their next decision. This dynamic approach, similarly to the loss aversion of the static approach in Prospect Theory, shows that mathematically fair investments with greater volatility may provide expected utility growth; therefore, investors maximizing the variance of the chosen fair investment – up to an optimal, risk-neutral state – may increase their utility gain for a fixed expected return. Hence, this is by definition risk-seeking behaviour, which can be seen in Figure 2 where we use the exponential utility function for CARA ($1 - e^{-ax}$). Here, we would like to add that later in Section 3.1, the results of our detailed empirical analysis on the VIX confirms that this risk-seeking behaviour is valid not only for tail risk but on the whole domain of losses, and therefore, it drives asset prices generally including tail risk calculations as well as the expected return. The calculation of the latter utility function is based on the equation

$$1 - e^{-5x} + 2.25\left(1 - e^{5(-0.05)}\right) = -2.25(1 - e^{5(-0.05)}) + 2.25(1 - e^{5(-0.05-0.05-x)}) \quad (1)$$

where *α = 5* is an average level of constant absolute risk aversion and *x = -5%* is the negative starting return indicating the reference point after a 5% loss. Based on the Monte Carlo approximation results for x, investors are risk-seeking in this case, until they are able to reach the x=7.18% return on the positive side. Here, we have to mention that the modified exponential function has a finite slope at zero, hence, not every (*α;x*) combination has a real solution for eq. (1). Less risk-averse investors (with low *α*) tend to become risk-seeking after huge declines in wealth as well but the maximal amount of loss (-x) inducing this phenomenon decreases as risk aversion increases.

**Figure 2: Fair investment after previous loss**

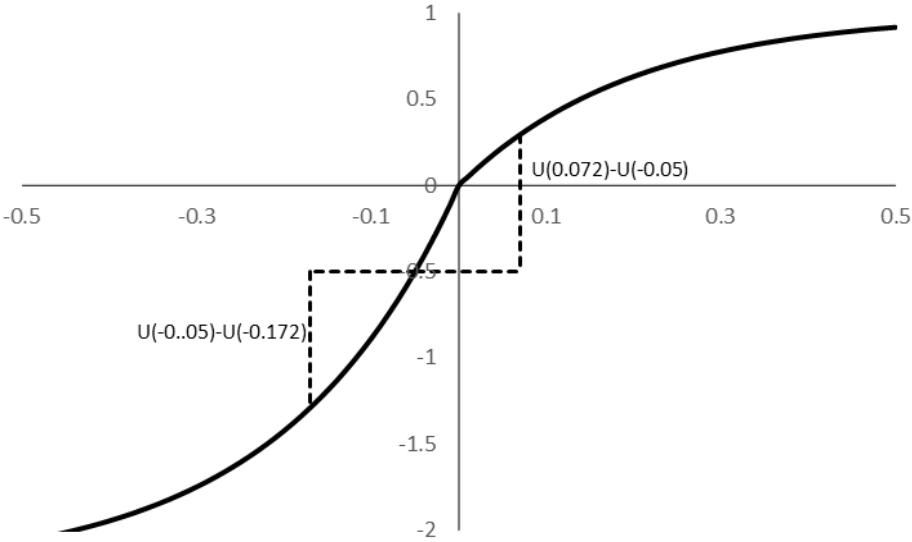

Notes: Figure 2 indicates the reference point change in dynamic utility perception subsequent to a loss. The horizontal and vertical axes stand for the wealth and value function changes respectively, where U indicates the utility function. The previous loss equals to -5%, whereas the mathematically fair investment with which investors become risk-neutral gives either 12.2% or -12.2% in addition to the –5% loss.

However, those who become risk-seeking maximize their risk (here the variance) for a given expected return until the utility growth due to the positive wealth change is greater than the utility loss due to negative wealth change. At this point they become indifferent to risk (i.e. the utility growth and loss of a further change in risk are equal). Depending on the initial portfolio this state differ and constitute together the risk-neutral curve that can be defined using the following method. We create a mathematically fair investment consisting of *(x + y)* gain and *(-x - y)* loss both with 50% probability. The reference point is *(-x)*. The investor keeps following risk-seeking behaviour until the utility growth and loss become equal, therefore:

$$U(-x) - U(-2x - y) = U(y) - U(-x) \quad (2)$$

By defining the utility on the negative side as a function of its positive equivalent, we can replace the utility for *x < 0* with *U(x) = - 2.25 U(-x)*, and we get:

$$-2.25U(x) + 2.25U(2x + y) = U(y) + 2.25U(x) \quad (3)$$

therefore

$$U(y) = -4.5U(x) + 2.25U(2x + y). \quad (4)$$

We can define an alternative version of eq. (4) by:

$$\frac{2.25[2U(x)+U(y)-U(2x+y)]}{1.25} = U(y) \quad (5)$$

One can see from this equation that due to the law of diminishing marginal utility *[2U(x)+U(y)-U(2x+y)]>0*, therefore, *U(y)* and *y* have to be positive in order to find a solution for the equation. The exact shape of the utility function can define the precise solution.

As for the variance of this investment, starting from the *-x* reference point choosing the mathematically fair investment with *(x+y)* amplitude, we can define:

$$\sigma^2 = \frac{[y-(-x)]^2 + [-2x-y-(-x)]^2}{2} = \frac{(y+x)^2 + (-x-y)^2}{2} = \frac{2(x+y)^2}{2} = (x+y)^2 \quad (6)$$

Furthermore, by assuming an alternative version of this investment with a positive expected return, we get similar results. Here, we keep the *-x* reference point but instead of *0,* an investment with an expected return of *0<$c_1$<x* is analyzed. In the followings we show that for an investment with the *$c_1$* expected return after *-x* loss (that is, having the *(-x+$c_1$)* reference point), the variance has to decrease in order to have the same utility change – in absolute terms – for both negative and positive wealth changes. The slope of this decrease can be defined by analyzing the precise utility function defined in eq. (5).

In order to find a solution for the equation between utility gains and losses we define the following situation. Given a mathematically fair investment with *$c_1$* expected return combined with the original reference point of *-x* the new reference point becomes *(-x+$c_1$)*. As mentioned above, the amplitude, having been *(x+y)* before, has to decrease by *-$c_2$* in order to keep the equality between utility differences caused by gains and by losses, therefore

$$U(y - c_2) - U(-x + c_1) = U(-x + c_1) - U(-2x + 2c_1 - y + c_2) \quad (7)$$

Converting the negative side to its positive equivalent again we get

$$U(y - c_2) = 2.25U(2x - 2c_1 + y - c_2) - 4.5U(x - c_1) \quad (8)$$

Through eq. (8), it is clear that in the case of *$c_1$=0* we get the same result as eq. (4) (thus *$c_2$=0*) and if *$c_1$=x*, then

$$U(y - c_2) = 2.25U(y - c_2) \quad (9)$$

Due to the multiplier 2.25, eq. (9) can be solved if and only if both sides are 0, therefore, $y=c_2$. According to eq. (4):

$$0 = 2.25[U(2(x - c_1) + y - c_2) - 2U(x - c_1) - U(y - c_2)] + 1.25U(y - c_2) \qquad (10)$$

We define the relation between $c_1$ and $c_2$ by using eq. (10) as a function. Assuming that the utility curve is a standard exponential function for constant absolute risk aversion that is $(1 - e^{-ax})$, we substitute the following: $x=c$, $y=d$ will be constant, while $c_1=x$ and $c_2=y$ will be the main variables. Eq. (10) is now rewritten:

$$2.25(1 - e^{-a(2c+d-2x-y)}) - 4.5(1 - e^{-a(c-x)}) - 1 + e^{-a(d-y)} \qquad (11)$$

The total derivative of this function by variable $x$ is:

$$\frac{d(2.25(1-e^{-a(2c+d-2x-y)}) - 4.5(1-e^{-a(c-x)}) - 1 + e^{-a(d-y)})}{dx} = -2.25e^{-a(2c+d-2x-y)}\left(\frac{da}{dx}(-(2c + d -$$

$$2x - y)) - a\left(2\frac{dc}{dx} + \frac{dd}{dx} - \frac{dy}{dx} - 2\right)\right) + 4.5e^{-a(c-x)}\left(-(c - x)\frac{da}{dx} - a\left(\frac{dc}{dx} - 1\right)\right) +$$

$$e^{-a(d-y)}\left(-(d - y)\frac{da}{dx} - a\left(\frac{dd}{dx} - \frac{dy}{dx}\right)\right) \qquad (12)$$

Since we know that according to eq. (10) this function has to be zero for all changes of $x$, its derivative also has to be zero. *df/dx* can be written as

$$\frac{df}{dx} = -2.25e^{-a(2c+d-2x-y)}\left(a\frac{dy}{dx} + 2a\right) + 4.5ae^{-a(c-x)} + a\frac{dy}{dx}e^{-a(d-y)} = 0 \qquad (13)$$

Thus, we can describe the relation between *x* and *y* as *dy/dx*:

$$\frac{dy}{dx} = \frac{4.5e^{-a(2c+d-2x-y)} - 4.5e^{-a(c-x)}}{e^{-a(d-y)} - 2.25e^{-a(2c+d-2x-y)}} \tag{14}$$

According to our Monte Carlo approximation for $(d-y)$ using $(c-x) = [0.01; 1]$ and $a = [0.5; 10]$ $\frac{d(d-y)}{d(c-x)}$, $\frac{d(d-y)}{d(c-x)} > 0$ for each solution, hence if the expected return *x* increases, *y* has to increase according to eq. (14) and thus the volatility of $(d-y) + (c-x)$ declines. In the case of small changes the same results apply for the power function used for CRRA as well.

Turning back to our original variable system *(x,y,c₁,c₂)*, we find that in the case of *0<c₁<x* this risk-seeking behaviour exists and the interval for variance-expected return combinations is known. Furthermore, the shape of the transition function is described by eq. (14).

Summarizing the above-mentioned derivation, we show that starting from the variance-expected return combination *(0;x)* (the minimum of the interval) and ending at the combination *((x+y)²;0)*, investors maximize the variance until they reach the "optimal" choice, the risk-neutral curve, which can be seen in Figure 3. We underline that this risk-neutral function is monotonically decreasing; therefore, according to eq. (14), as the expected return of the investment *(c₁)* increases, the variance *(c₂)* has to decrease. Hence, instead of investment *A,* an investor would choose investment *A'* with the same expected return and higher variance, which is completely irrational in standard equilibrium models. The milestone paper of Thaler and Johnson (1990) also confirms these results where we see a good example of investors becoming risk-seeking subsequent to losses if breaking even is possible, which latter is practically always valid in the case of capital markets.

**Figure 3: Risk-seeking until risk neutrality**

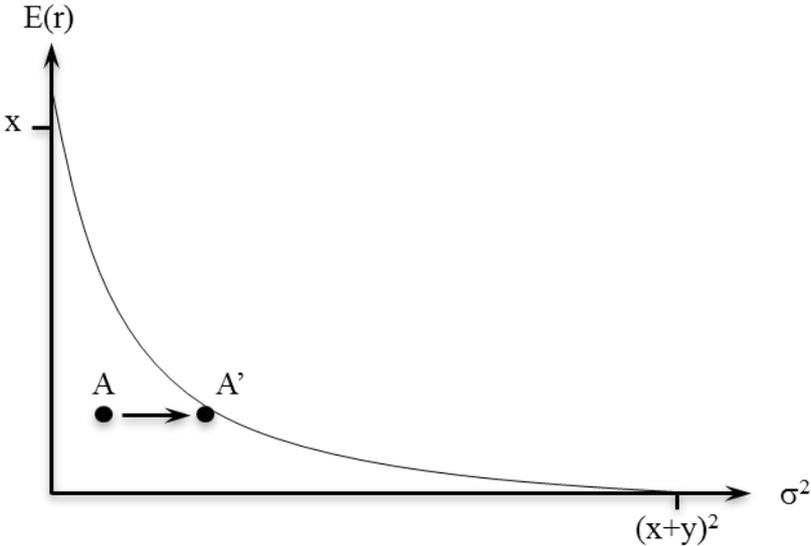

Notes: Figure 3 represents the optimization of risk-seeking investors. They maximize utility by increasing risk (thus decreasing EDR) until they reach the risk neutral point A'. Depending on the initial portfolio choice alternatives of A' constitute the risk-neutral curve.

As our result is derived in a variance-expected return system, it can be implemented in standard equilibrium models as we show in the next section. However, standard asset pricing models require symmetrically distributed returns, which is clearly not the case in the real markets. Therefore, in Section 3 we propose a model applying Expected Downside Risk (*EDR*) as risk measure, which allows us to describe the same situation with an alternative approach that is able to manage any type of return distribution.

## 2.3 Implementation in standard models

In order to place the above-mentioned behaviour in asset pricing theories we combine the results of Expected Utility Theory (EUT) and Prospect Theory (PT). These two theories – although based on different assumptions – have many similar properties. Although EUT examines the utility of total wealth and PT explains the change of utility for gains and losses, both utility functions are concave and strictly monotonically increase on the positive side. We assume that the utility function described by PT is actually the same as the EUT function from the reference wealth $W = 1$. Therefore, the behaviour of completely rational investors, who are not influenced by different heuristics (such as the anchoring effect) and are perfectly informed, can be described by both EUT and PT. These market participants have risk-averse behaviour; therefore, the base model of CAPM, Modern Portfolio Theory (MPT) (Markowitz, 1959), can sufficiently explain their preferences.

However, risk-seeking behaviour can also be implemented in MPT using the following method. We have already shown that risk-seeking always has a limit where it reaches risk neutrality and this converges to zero variance with the growth of expected return (since it decreases monotonically in a variance-expected return system). This risk-neutral curve (RNC) consists of points where investors are risk-neutral, where their utility depends only on the expected return of investments and it is not influenced by risk (here variance). Therefore, investors choose the portfolio with the highest possible return on their RNC. As this curve crosses the MPT defined curve of efficient portfolios, the efficiency frontier (EF) – by definition the portfolio with the highest expected return at a given risk level – also the optimal choice – is exactly the intersection of these two curves illustrated in Figure 4. Thus, risk-seeking behaviour has no effect on the efficient portfolios of standard equilibrium models; therefore, it can be implemented in asset pricing regressions. However, we have to take into consideration that it may change the overall volatility of markets if the efficiency frontier is an endogenous variable as well (which we show in Section 5).

**Figure 4: Implementing risk-seeking**

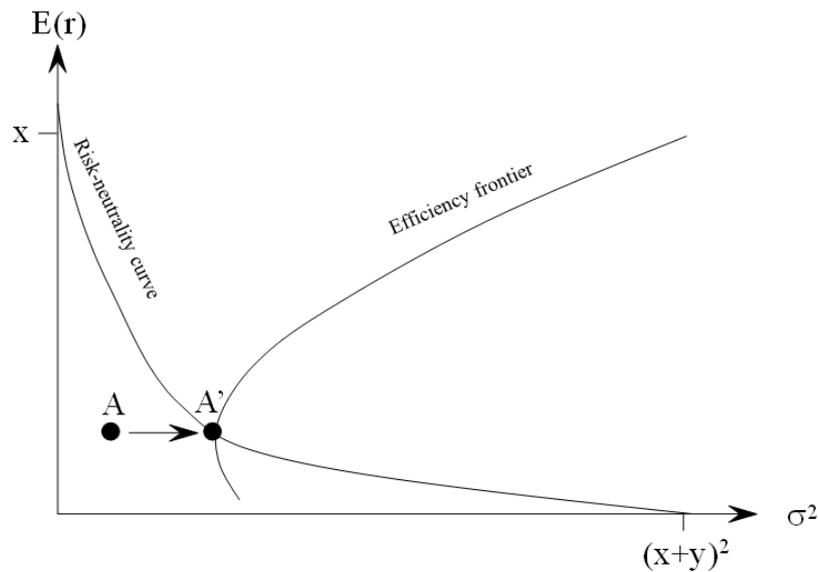

Notes: Figure 4 shows the optimization of risk-seeking investors. They maximize utility by increasing risk (thus increasing variance) until they reach the risk neutral point A' then, as the only determinant of utility is the expected return on the risk-neutral curve, they pick the highest reachable portfolio, which turns out to be exactly the cross-section of the risk-neutrality curve and the efficiency frontier.

## 2.4 Evidence of risk-seeking behaviour in empirical data

The existence of risk-seeking behaviour in everyday life is shown in examples of Shefrin (2002) and Thaler and Johnson (1990). Furthermore, we provide some evidence as well in the following.

We analyze daily returns of the Standard and Poor's 500 Index from 1950 to 2013 and find strong evidence that supports our results of risk-seeking behaviour described above. Regardless of the time, the negative return in period *t* causes higher volatility in *t+1*. We divide our experiment into monthly and weekly analysis. We assume that two distinct effects generate the volatility change in period t+1. First, a massive gain or loss in *t* affects next period volatility due to the behaviour shown above (Black, 1976; Bollerslev, 1986; Engle, 1982; Christensen et al., 2015; Zhou and Nicholson, 2015); second, the gains or losses in interval *t+1* also influence volatility due to equilibrium pricing. With this method we can rule out the possibility that the volatility change is only due to the price movements in the same period or vice versa. We find that extreme previous period return has a greater effect on current volatility than the return in the same period.

In our analysis, massive loss is defined as return below the 10% percentile and massive gain as return above the 90% percentile of the density function. We find that in the case of a massive weekly loss, the increasing effect on next period volatility is higher than in the case of monthly loss; however, both are significant. We argue that this effect is due to the availability heuristic – which drives investors to rely more on recent information – which causes previous weekly returns to have a stronger influence than previous monthly returns.

Table 1 shows that after a monthly price fall, the results are only significant if the next month has a negative return (thus the S&P 500 is in a downtrend); however, the aggregate monthly results are also significant. In the case of a weekly price fall, regardless of the trend over the next week, volatility increases significantly.

**Table 1: Effect of price jumps and price falls on volatility**

|  | average | p-value |
|---|---|---|
| Volatility change after price jump in monthly uptrend | -0.65% | 0.4316 |
| Volatility change after price jump in monthly downtrend | 3.36% | 0.3294 |
| Volatility change after price fall in monthly uptrend | -6.99% | 0.0663 |
| Volatility change after price fall in monthly downtrend | 33.91% | 0.0001*** |
| Volatility change after price jump in weekly uptrend | -2.61% | 0.2158 |
| Volatility change after price jump in weekly downtrend | 9.44% | 0.0502 |
| Volatility change after price fall in weekly uptrend | 26.76% | 0.0000*** |
| Volatility change after price fall in weekly downtrend | 37.03% | 0.0000*** |
| Avg vol change after price jump monthly | 0.61% | 0.4309 |
| Avg vol change after price jump weekly | 1.98% | 0.2550 |
| Avg vol change after price fall monthly | 12.10% | 0.0078** |
| Avg vol change after price fall weekly | 31.38% | 0.0000*** |

Notes: We used the daily historical data from the S&P 500 Index for the period of January 2, 1950 to May 6, 2013. We define price falls as a monthly (or weekly) return below the 10$^{th}$ percentile of the monthly (or weekly) return distribution and price jumps as monthly (or weekly) returns above the 90$^{th}$ percentile of the monthly (or weekly) return distribution. Uptrend means that the next period has a positive return, while downtrend stands for a negative return. Monthly (or weekly) volatilities are based on daily returns and their change is calculated using logarithmic return. The averages presented are the mean weekly and monthly volatility changes over the previous period in the indicated sub-cases and the p-values stand for the respective Student's t-test for them with the null of the average being zero.

According to our model, this phenomenon can be explained by the risk-seeking effect after losses. Since the Standard and Poor's 500 Index is a value-weighted index, it reflects the changes in asset allocation. In the case of price falls, investors suffer from losses and turn to riskier investments. They substitute their less risky portfolio elements with riskier ones, thus the increased demand for risky investments elevates their price and market capitalization, while the opposite happens for less risky ones. This way the weights of riskier investments in the S&P 500 increase; therefore, their price movements and volatility will contribute to the index more, increasing the volatility of the S&P 500 itself as well. These results are confirmed by literature on mutual fund activity as well, according to which a negative relationship was found between returns and subsequent money inflows (Warther, 1995; Goetzman and Massa, 1999; Edelen and Warner, 2001) and between contemporaneous inflow of equity and bond funds (Goetzmann et al., 2000).

## 3. EXPECTED DOWNSIDE RISK (*EDR*) AS RISK MEASURE

Since the introduction of the expected return-variance relationship in MPT and the beta in the CAPM, numerous attempts have been made to offer an alternative risk measure. Amongst the most important ones we find Value-at-Risk (Holton, 2003), which gained popularity amongst quantitative financial professionals in the late '80s (mainly due to the crisis in 1987). This measure provides an entirely new method to calculate the true risk of derivatives, options and other nonstandard financial assets as it shows the maximal value that investors can lose with a predefined confidence interval (or probability), which is based on the true density function of these investments. Although it exhibits many advantages over volatility and beta, it misses the important characteristics of subadditivity and convexity (Rockafellar and Uryasev, 2000). Conditional Value-at-Risk (CVaR) is introduced to solve the problems of subadditivity and convexity, which makes CVaR one of the most precise risk-measuring techniques today (Krokhmal et al., 2002). However, the predefined significance level in CVaR also exhibits some disadvantages in quantitative methods. Choosing the optimal level of $\alpha$ is difficult and influences heavily the investment choice in modeling. This problem with the subjective confidence level has been questioned in Huang et al. (2012) as well, however, their proposed risk-measure, the extreme downside risk comes with two possible flaws: first, it is based on extreme value theory, therefore, its fitted distribution may not represent the true behaviour of asset prices; second, its left tail index is difficult to interpret for the average investor, and therefore, it may not account for the true premium required. The performance measure Omega (Keating and Shadwick, 2002) avoids the subjectivity problem by measuring the ratio of the expected abnormal return below and above zero; however, its use in equilibrium modeling is fairly moderate since investors are rather interested in downside risk and expected return. Alternative measures similar to ours are semi-variance (Markowitz, 1968) and its standardized version, the downside beta (Estrada, 2007), which consider only the downside-risk of returns, however, these measures cannot be fully implemented in equilibrium asset pricing under the assumption of expected utility theory and again their relevance in the risk premium is questionable due to their complexity. We apply Expected Downside Risk as risk measure with which we can eliminate the above-mentioned disadvantages.

Our definition of *EDR* is based on VaR and CvaR, which are discussed in Appendix 1 in detail. Similar to Omega, semi-variance and downside beta, *EDR* does not apply any subjectively fixed *α* percentile to measure tail risk but is defined as a risk measure that covers the loss function on the whole domain, although, in contrast to Omega, we examine losses relative to the expected return *(E(r))* instead of zero. In other words, we define Expected Downside Risk as the negative Conditional Value-at-Risk for *α* significance level, where *VaR$_α$(x)* is the expected return. This is shown in eq. (15), where the *EDR* of investment *x* is described:

$$EDR(x) = p(r(x) \leq E(x))^{-1} \int_{r(x,y) \leq E(r(x))} r(x,y)p(y)dy, \qquad (15)$$

where $r(x,y)$ stands for the return distribution of portfolio x, p(y) for the probability density function and E(x) for the expected return. We illustrate the *EDR-E(r)* system in Figure 5. Since *EDR* measures the whole downside risk and considers return above zero as well, we may discover some investment opportunities that reflect well the positive time preference of investors. This phenomenon is due to the fact that consumption in the present causes higher utility growth for investors than the same consumption in the future; therefore, they require compensation in the future in exchange for lending money (thus utility) in the present. Hence, investments may exist where *EDR(x)* exceeds *0* because their risk is so low that the average of negative returns in absolute meaning is smaller than the expected return itself. Even if the investment turns out to have a lower return than the expected one, the Expected Downside Risk is positive; that is, the investor realizes a positive expected return even if the return falls short of the expectation. This can be interpreted as the risk-free return defined in standard equilibrium asset pricing models; however, in reality, none of them are entirely risk-free. Therefore *EDR(x)* is always lower than *E(r)*. This can be illustrated by the line with *s=1* slope in the *EDR-E(r)* coordinate system. None of the portfolios can reach the area below this line.

We illustrate the system of *{EDR(x),E(r)}* pairs in Figure 5 to the right. The weights and the values of the pairs are calculated by Monte Carlo simulation with the following data and parameters. We use historical, annual returns from 1993 to 2014 of 22 different assets: 20 randomly sampled shares, the 3-month Treasury bill and the value-weighted equity index of the Center for Research in Security Prices (CRSP). During the simulation 10,000 randomly weighted portfolios are generated and for every portfolio we calculate the average (expected) annual return, variance and the *EDR*. In Figure 5 the horizontal axis shows the Expected Downside Risk, while the vertical axis indicates the expected return. The efficiency frontier (the highest expected return for every *EDR* value) is very similar to the one in the MPT (to the left in Figure 5) since it is concave, although its slope is decreasing instead of increasing. So, the optimization with indifference function (which is convex in MPT) seems to have a unique solution for *EDR* equilibrium as well.

**Figure 5: The Markowitz (left) and the *EDR-E(r)* system (right)**

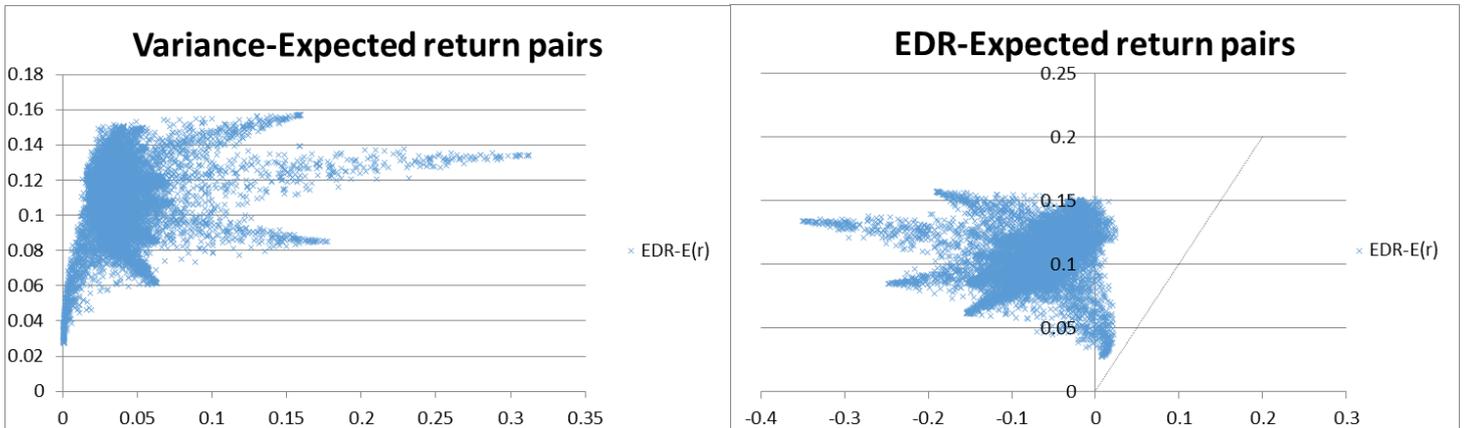

Notes: Figure 5 indicates indicates simualted variance - expected return (left) and EDR - expected return (right) pairs. The simulated portfolios contain 10000 randomly weighted portfolios of 22 different assets from 1993 to 2014: 20 randomly sampled individual shares, the value-weighted CRSP equity index and the T-bill with three-month maturity.

In Figure 6 we show the difference between applying normal or Gaussian PDF and using historical distribution of the returns in *EDR* calculation to underline the importance of empirical risk measures. For investments with low expected return, historical distribution seems to have a lower risk than the estimation with normal PDF, while for those with high expected return the risk is higher using empirical data, the latter of which reflects the fat-tail distribution of returns. The former phenomenon is due to the fact that less risky investments (e.g. government bonds) have highly skewed distribution, thus their expected return is well above the median; therefore, the loss integrating interval to *E(r)* is wider than in the case of normal distribution, where it would be exactly the median. The Student's t-test with the null hypothesis of the difference being zero is rejected at extremely low p-values (less than 0.0000), hence, applying the historical distribution indeed matters.

**Figure 6: The effect of true distribution relative to normal**

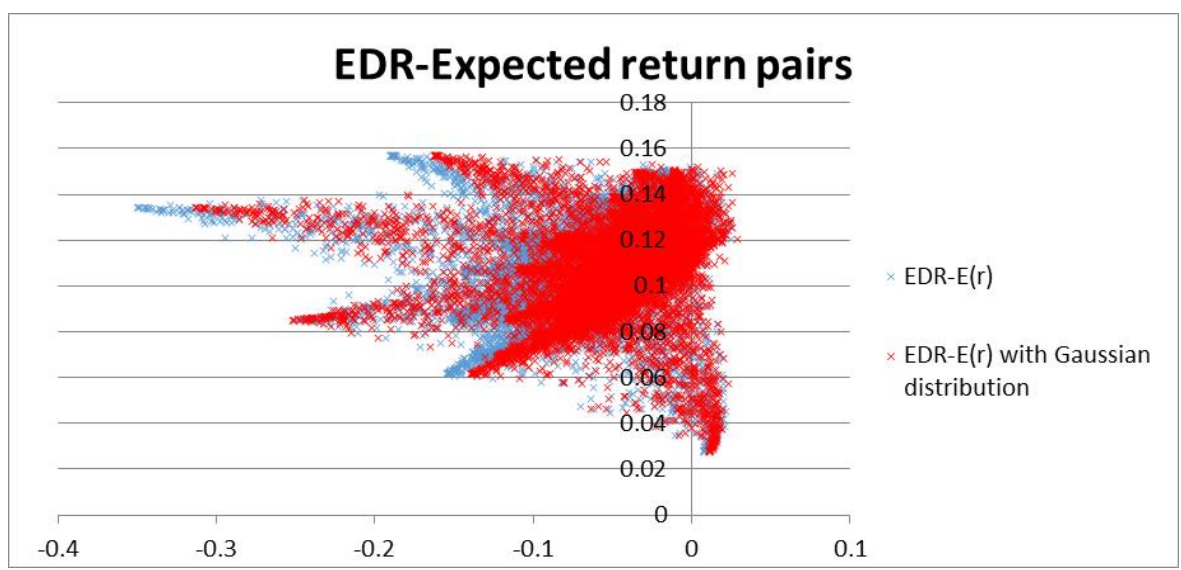

Notes: Figure 6 indicates EDR (horizontal axis) and E(r) (vertical axis) pairs using Gaussian and historical distribution of yearly returns of 10000 randomly weighted portfolios containing the annual returns of 22 different assets from 1993 to 2014: 20 randomly sampled individual shares, the value-weighted CRSP equity index and the T-bill with three-month maturity.

Although numerous techniques exist to solve the minimization problem for $CVaR_\alpha(x)$ (e.g. Krokhmal et al., 2002), which can easily be applied to $EDR(x)$ as well, our main goal is not to describe these methods or the calculated efficiency frontier but to combine them with the utility measure in Prospect Theory, hence, to describe the optimal choice of investors with precise quantitative analysis.

### 3.1. Testing the probability level of *EDR*

We assume that the Volatility Index of S&P 500, the VIX, also commonly known as the "fear index," reflects well investors' reaction to price changes. If the VIX changes much, it means that investors' reaction to the given price movement is fairly strong, or they are influenced heavily by the given change in prices.

We examine the price changes of the S&P 500 Index from 1990 to 2013 (using the same data as in 2.4), and then compare this to the changes of the VIX Index in the same period. Our results confirm that changes in *EDR* are a really important factor for investors; thus, the risk premium correlates strongly with our risk measure. The results show that the 5% α – usually chosen for Conditional Value-at-Risk – is not appropriate; however, using higher levels of α produces higher predictive power to describe investors' decisions.

In our test we examine the changes of the VIX for every α percentile of the distribution of S&P 500 daily returns. We apply a *t-test* to examine whether the changes of the VIX Index are significant. The p-values of the changes are presented in Figure 7.

**Figure 7: Student-t values for changes of VIX**

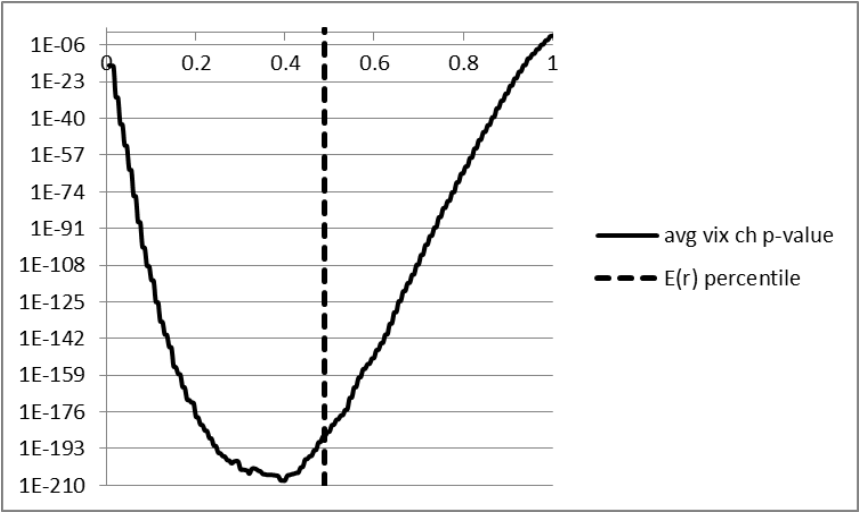

Notes: The graph above indicates the p-values of Student's t-tests with the null hypothesis that contemporaneous changes of VIX conditional on quantiles of S&P500 returns indicated on the horizontal axis are equal to zero.

According to the left side of the curve, increasing the *α* percentile initially raises the prediction power, thus returns higher than the 5% percentile of the distribution also play an important role in investors' decision. Although we find the *p-value* reaching its minimum at the 40% percentile where it is $1.96 \cdot 10^{-208}$, raising α further to the *E(r)* percentile (at 49%), which we also use for *EDR*, we get $8.1 \cdot 10^{-188}$ for the *p-value*, which is clearly not far from its minimum and is very significant. Therefore, we may state that the interval of the return distribution used for *EDR* calculation, hence *EDR* itself too, does have an effect on investors' decision and seems to be a good risk measure. Here, we add that comparing p-values of such magnitude may seem irrelevant to the reader, however, these extermely low values are due to the high number of observations. Therefore, we argue that the trends are much more relevant than the magnitudes themselves. Optimization and evidence for picking a given probability level for our risk measure requires comparing such numbers.

In addition to the aforementioned choice of the applied cutoff level for expected loss we provide in the followings a description how *EDR* affects asset prices and expected return and a brief empirical test whether it is able to better capture the expected return than its alternatives.

### 3.2. *EDR*-based asset pricing

As we mentioned before, the level of risk aversion of an investor can be defined by a unique risk aversion parameter "a" (Pratt, 1964). Since we assume that the Kahneman and Tversky utility function has the same convexity on the right side as the expected utility function in EUT, the behaviour of an individual can be described in both risk-averse and risk-seeking cases with the help of this measure. Therefore, our model is able to define the optimal choice for every investor with the constraints that the goal is utility maximization and the efficiency frontier is known.

In the case of risk aversion, the approximation of expected utility could be used in our model as well, that is

$$U(F) \cong E(F) - 0.5a\sigma^2 \tag{16}$$

First, we use normal distribution as approximation, which allows for tighter conditions, however, true distributions can be calculated in the same way. This way we can define *EDR* as the function of expected return and variance:[1]

$$EDR(x) = E(r_x) - 0.8\sigma \tag{17}$$

According to eq. (17), we can substitute the volatility ($\sigma$) with the Expected Downside Risk (*EDR*); therefore, the approximating function (16) can be implemented in the *EDR-E(r)* system:

$$U = E(r_x) - \frac{0.5}{0.8^2} a[E(r_x) - EDR(x)]^2. \tag{18}$$

We can define the slope of the iso-utility function in the *EDR-E(r)* system as the total derivative should be zero, hence

$$\frac{dU}{dEDR(x)} = \frac{a}{0.8^2}\big(E(r_x) - EDR(x)\big) + \frac{dE(r_x)}{dEDR(x)}\left(1 - \frac{a}{0.8^2}\big(E(r_x) - EDR(x)\big)\right) = 0. \tag{19}$$

Therefore the sensitivity of the expected return for the change of the Expected Downside Risk is

$$\frac{dE(r_x)}{dEDR(x)} = -\frac{\frac{a}{0.8^2}\big(E(r_x)-EDR(x)\big)}{1-\frac{a}{0.8^2}\big(E(r_x)-EDR(x)\big)} = 1 - \frac{1}{1-\frac{a}{0.8^2}\big(E(r_x)-EDR(x)\big)}. \tag{20}$$

---

[1] in the case of normal distribution EDR=CVaR$_{0.5}$= $\int^{E(r)} r \cdot \left(\frac{1}{\sigma\sqrt{2\pi}} e^{\frac{[r-E(r)]^2}{2\sigma^2}}\right) dr$

In order to determine the concavity of the iso-utility function we use the second derivative, that is

$$\frac{d^2 E(r_x)}{dEDR(x)^2} = \frac{\frac{a}{0.8^2}}{\left(1-\frac{a}{0.8^2}\left(E(r_x)-EDR(x)\right)\right)^2}. \qquad (21)$$

This is always positive, hence, iso-utility functions are convex. We distinguish three cases, which are $EDR(x) < E(r_x) - \frac{0.8^2}{a}$, $EDR(x) = E(r_x) - \frac{0.8^2}{a}$, and $EDR(x) > E(r_x) - \frac{0.8^2}{a}$. For the first case we have

$$1 - \frac{a}{0.8^2}\left(E(r_x) - EDR(x)\right) < 0, \frac{dE(r_x)}{dEDR(x)} > 1. \qquad (22)$$

For the second case $\frac{dE(r_x)}{dEDR(x)}$ is not differentiable, and the third case implies

$$1 - \frac{a}{0.8^2}\left(E(r_x) - EDR(x)\right) > 0, \frac{dE(r_x)}{dEDR(x)} < 0. \qquad (23)$$

Except for extremely high risk aversion combined with very high expected return, which is highly unlikely as described in Appendix 2, the third case applies for real situations practically always. Therefore, iso-utility functions have negative slope and are convex, hence, optimization methods using the positive sloped and concave efficient frontier can determine the optimal choice in the *EDR-E(r)* system as well.

## 4. RISK-SEEKING BEHAVIOUR IN EDR ENVIRONMENT

Standard asset pricing models based on the assumption of normally distributed returns (which overestimate the effect of diversification sometimes and do not consider the autoregression of short-term returns) cannot describe precisely risk-seeking behaviour when return distribution is asymmetric and fat-tailed. As we have mentioned before, the variance-based theory is applicable only in the case of symmetrically distributed returns. Therefore, we introduce a parameter that can describe the risk of an investment without having to consider its skewness or kurtosis, one that can define the risk on both the negative and positive side for any type of distribution. Using Expected Downside Risk we get the optimal solution for this problem.

In order to describe the *EDR*-based method, first we introduce a *Prospect* variable *[Pr(x)]*. *[Pr(x)]* is the opposite of Expected Downside Risk; thus, it measures the expected value of the outcomes higher than the expected return. Since

$$\alpha EDR(x) + (1-\alpha)Pr(x) = E(r_x) \tag{24}$$

where $\alpha$ is the probability that return will be below the expectation, so $\alpha = p(r(x) \leq E(r(x))$; therefore, *Prospect* is defined by:

$$Pr(x) = \frac{1}{(1-\alpha)}[E(r_x) - \alpha EDR(x)] \tag{25}$$

Modifying Figure 2 with the *Prospect* we get Figure 8.

# Figure 8: Risk-seeking with Prospect

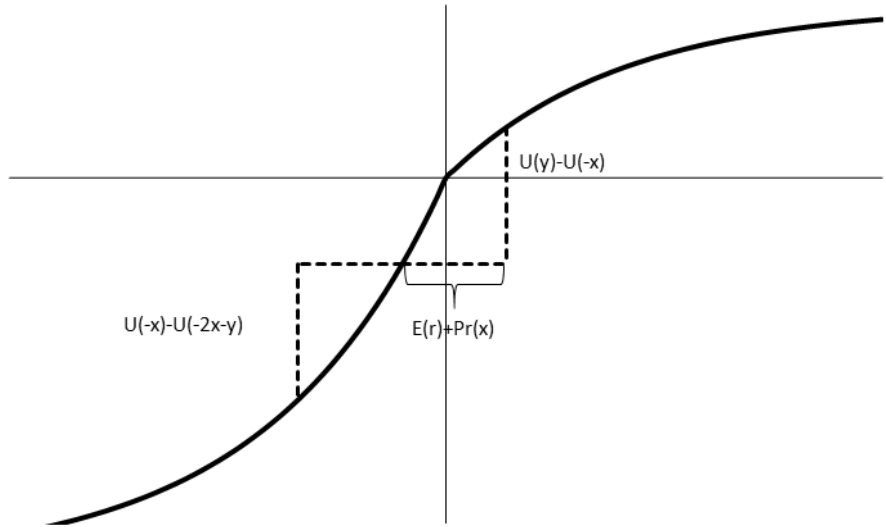

Notes: Figure 8 represents the reference point change in dynamic utility perception subsequent to a loss. The horizontal and vertical axes stand for the wealth and value function changes respectively. U indicates the utility function, E(r) the expected return and Pr the prospect. The previous loss equals to –x, whereas the mathematically fair investment gives either +x+y or –x-y in addition to the –x loss.

This way we can omit the symmetry requirement of distributions since the *α*-weighted average of the *Prospect* and the *Expected Downside Risk* always sums up the expected return. Therefore, we define every investment as fair investment by simplifying the distribution to these two outcomes. So, we omit the main assumption of modeling with variance, the requirement of symmetric distribution of returns.

Figure 9 clearly shows the similarity of risk-seeking in the models based on variance and on *Prospect.* In fact, the only difference is that in the case of using *Prospect* the model has a solution for asymmetric discrete and continuous distributions as well, and according to eq. (14) the change of expected return causes an opposite change in the value of *Prospect.* In order to capture the relation we repeat eq. (14):

$$\frac{dy}{dx} = \frac{4.5e^{-a(2c+d-2x-y)} - 4.5e^{-a(c-x)}}{e^{-a(d-y)} - 2.25e^{-a(2c+d-2x-y)}} \qquad (14)$$

If we substitute *(d-y-(-c+x))* in eq. (14) (which measures the distance between the expected value of positive outcomes and the reference point) with $Pr(x)$, we are able to implement this situation in the *Pr-E(r)* system. Based on this substitution, *y* decreases as c decreases, $Pr(x) = d + c - y - x$ has to decrease for the increase of *c* and *y* values since both variables have a negative effect on it. Therefore, we can define a risk-neutral curve similarly to the one in the variance-expected return system, which is shown in Figure 9. The minimum and maximum points of the interval are *Pr=y-(-x)=x+y* for *E(r)=0* and *Pr=0* for *E(r)=x*, which is the same as in the model based on variance. In fact, *Pr* could take on values below *0*; the only restriction is that it has to be above the expected return. Although, in reality, we do not deal with investments like this, this case is included in the *EDR-E(r)* model as well.

**Figure 9: Risk-seeking in *Pr-E(r)* environment**

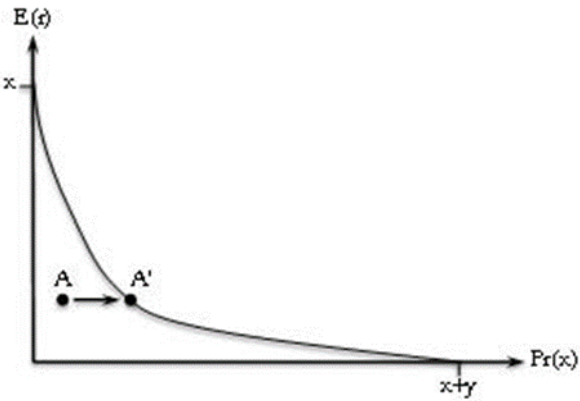

Notes: Figure 9 shows the optimization of risk-seeking investors. They maximize utility by increasing risk (thus decreasing EDR) untill they reach the risk neutral point A' then, as the only determinant of utility is the expected return on the risk-neutral curve

Although our model is based on *Expected Downside Risk*, not *Prospect*, we have already defined the relation between the two in eq. (25). Substituting this to eq. (14) (where *E(r)=x* and *Pr(x)=d+c-y-x*) we get

$$EDR(x) = \frac{x-(1-\alpha)(d+c-y-x)}{\alpha} = \alpha^{-1}\big((2-a)x + (1-a)y - (1-a)c - (1-a)d\big). \quad (26)$$

Since *(1-α)* is positive, one can see that the growth of both x (the expected return) and y increases *EDR(x)*; therefore, the risk-seeking phenomenon can be implemented in our *EDR-E(r)* model in the way illustrated in Figure 10, where the risk-neutral curve is presented.

**Figure 10: Risk-seeking in *EDR-E(r)* environment**

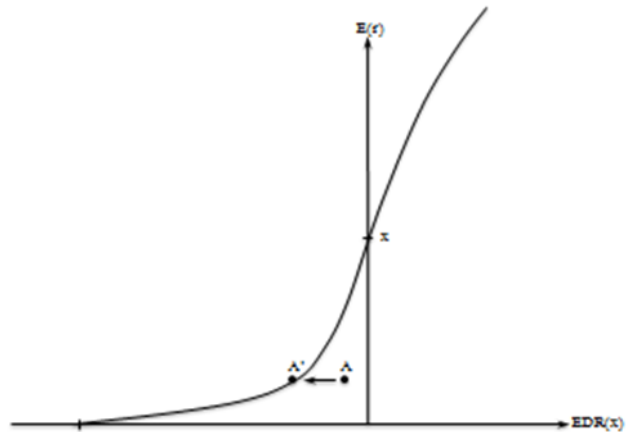

Notes: Figure 10 shows the optimization of risk-seeking investors. They maximize utility by increasing risk (thus decreasing EDR) until they reach the risk neutral point A'. Depending on the initial portfolio choice alternatives of A' constitute the risk-neutral curve.

Figure 10 shows clearly that the efficiency frontier of the *EDR-E(r)* model and the risk-neutral curve, similarly to the variance-based regression, produce a unique optimum again. By accepting that this risk-seeking behaviour exists, investors maximize the risk according to eq. (26), and thus minimize the *EDR* of their investment for every *E(r)* expected return up to the frontier curve. At this point they become risk-neutral; thus, their optimal choice depends only on *E(r)*. Therefore, they choose the point with the highest expected return on their risk-neutral curve (RNC), which is an efficient portfolio, precisely the intersection of the efficiency frontier of the *EDR-E(r)* model and the RNC as presented in Figure 11.

**Figure 11: Risk-neutral curve and efficiency frontier**

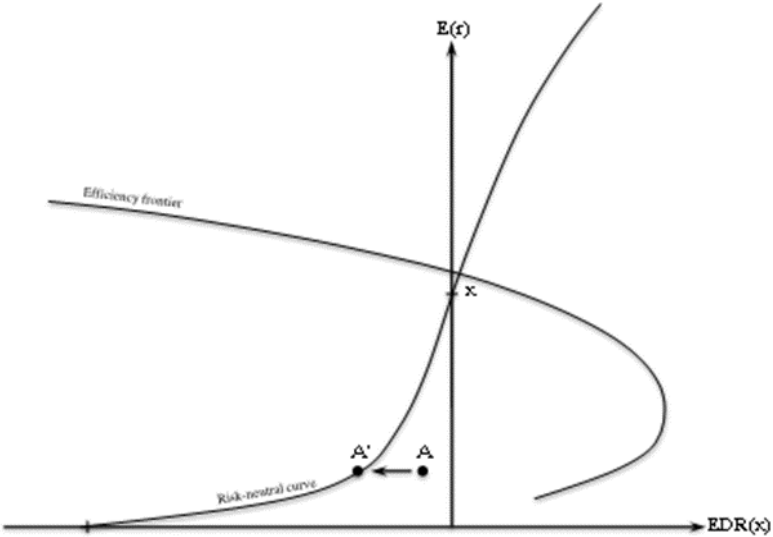

Notes: Figure 11 shows the optimization of risk-seeking investors. They maximize utility by increasing risk (thus decreasing EDR) until they reach the risk neutral point A' then, as the only determinant of utility is the expected return on the risk-neutral curve, they pick the highest reachable portfolio, which turns out to be exactly the cross-section of the risk-neutral curve and the efficiency frontier.

Hence, our model based on Expected Downside Risk is applicable for risk seekers as well and thus we do not have to deny the existence of these investors. If this behaviour is limited (and is as shown above), we can implement the phenomenon, and the calculation based on the efficient frontier stays intact, the expected return for every *EDR* will be the same as in the case without risk-seeking.

## 5. THE EFFECT OF LIMITED BORROWINGS

In this section we omit one of the main, although very unrealistic, assumptions of standard asset pricing models, unlimited borrowings. CAPM defines the capital market line (CML) as the set of efficient investment opportunities, including risk-free and risky assets, that goes to infinity. However, this assumption is fairly unrealistic (Holmes et al, 2015). In most of the cases there is no opportunity to invest in such positions. Using realistic factors the expected return of portfolios is completely different from that of standard regressions. We use the finite borrowing constraint that is available at different interest rates. In order to create a leveraged model we create the following assumptions, which are more realistic than the ones used in standard models (Black, 1972; Fama and French, 2004):

- For each investor borrowing is limited (as it is either investment credit or proceeds from short sales). In the following, this limit is measured by *(1+x)* (for example, in the case of 2:1 leverage *x=1*).
- Every time investors use leverage the lending institution defines a margin that involves automatic closing of the position or liquidation if the value of the portfolio reaches its *m* percentage. Analytically this means that the return reaches a *(-1+m)* loss; therefore, *r=(-1+m)*.

Due to these assumptions, investors gain other advantages in exchange for paying the interest rate. On the one hand, in the vast majority of contracts, they get insurance "for free" due to liquidation at the margin call. In this case, the investors cannot lose more than their own invested money; however, it would be possible through a leveraged portfolio without marginal requirements. This reduction of risk has no excess cost for them; however, they get some of the negative risk eliminated, and thus get a higher expected return, as is described in Figure 12 and in the following analytics.

**Figure 12: Effect of leverage with marginal requirements**

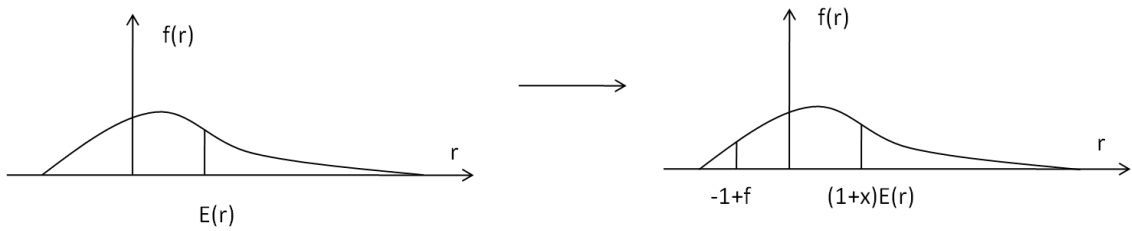

Notes: The graphs represents the effect of leverage and margin call: as the leverage increases to (1+x) the expected return would increase with it, however, the margin call protects both the investor and the lender from losing more than the cutoff level (-1+f), therefore, further increasing the expected return.

$$E(r)_L = E(r)_P(1+x) - r_c x - p(r_Q < -1+m)CVaR_{Q,p(r_Q<-1+m)} - p(r_Q < -1+m)(-1+m) \tag{27}$$

$$E(r)_L = E(r)_P(1+x) - r_c x - p(r_Q < -1+m) \cdot \left(CVaR_{Q,p(r_Q<-1+m)} - VaR_{Q,p(r_Q<-1+m)}\right) \tag{28}$$

where:

- $E(r)_L$ is the expected return of L leveraged portfolio with margin requirements
- $E(r)_P$ is the expected return of P unleveraged portfolio
- $(1+x)$ is the leverage
- $r_c x$ is the interest rate for borrowing multiplied by the borrowed quantity (the total cost of borrowing)
- $p(r_Q < -1+m)$ is the probability of Q leveraged portfolio without margin requirements generating a return below (-1+m)
- $CVaR_{Q,p(r_Q<-1+m)}$ is the Conditional Value-at-Risk of Q portfolio at p probability
- $VaR_{Q,p(r_Q<-1+m)})$ is the Value-at-Risk of Q portfolio at p probability

Since *CVaR≤VaR* is always true for fixed distribution and probability and $r_Q$ is multiplied by *(1+x)* for *x* leverage, we get increasing marginal expected return in the case of margin requirements instead of constant marginal expected return. So, the relation between the leverage of the portfolio and the expected return is not linear, and the function describing *EDR-E(r)* leveraged portfolios $\frac{dE(r)}{dx}$ is not constant.

According to this deduction one can create the following leveraged position. If *A* and *B* are unleveraged portfolios, *EDR$_A$=EDR$_B$* and *E(r)$_A$>E(r)$_B$*, A is "more efficient"; therefore, it is the optimal choice. However, the reduction effect of margin requirements can have the opposite result for leveraged expected returns if *A* and *B* have different probability distributions. This causes *E(r)$_{L;A}$<E(r)$_{L;B}$* if the return distribution of portfolio *B* has fatter tails.

$$p(r_{A(1+x)} < -1 + m)(CVaR_{A(1+x),p(rA<-1+m)} - VaR_{A(1+x),p(rA<-1+m)}) - p(r_{B(1+x)} < -1 + m)(CVaR_{B(1+x),p(rB<-1+m)} - VaR_{B(1+x),p(rB<-1+m)}) > [E(r)_A - E(r)_B](1 + x)$$

(29)

where *r$_{B(1+x)}$* is the leveraged expected return of portfolio *B* using *(1+x)* leverage and no margin requirements. The situation mentioned above is presented in Figure 13 where *A* and *B* are portfolios without leverage, *A'* and *B'* are portfolios with *(1+x)* leverage, *D* is the interest rate paid for borrowing and the 45° dashed line indicates the frontier that no portfolio can exist below since *EDR≤ E(r)*.

**Figure 13: Effect of different probability distributions on portfolios**

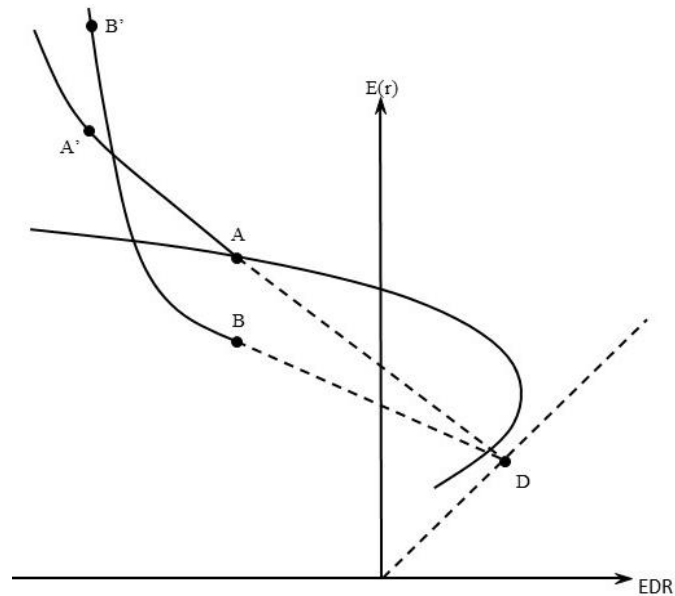

Notes: The graph represents two portfolios with different return distribtuions. Due to the limited loss (margin call), a portfolio with higher expected loss below the cutoff value could yield better leveraged opportunity while not being efficient without leverage.

According to Figure 13, rational investors having risk-averse behaviour may hold portfolios that generate less expected return for a given risk with unleveraged conditions. Therefore, we accept the fact that investors hold positions that seem to be "inefficient" initially may be a rational choice. Furthermore, this phenomenon also contradicts the strict dominance of diversification mentioned in standard asset pricing theories. While investors can reduce the volatility to a point through diversifying their portfolios, the distribution of their returns tends to converge to normal distribution as the number of investments grows. Since normally distributed portfolios have much less probability at the tails, the positive effect of margin requirements is also decreased. Therefore, it is not enough to sacrifice everything to diversify without considering any other parameters; one has to analyze the optimal choice in regard to the effect of leverage and the liquation at the margin call.

These parameters can take on values from a wide range; however, with the necessary information, the regression is very precise. In order to describe the behaviour of "investors we use," the "a" Arrow-Pratt measure of risk aversion (CARA) can be defined through various methods such as questionnaires and observation tests (Barsky et al., 1997; Hanna and Lindamood, 2004). Information technology today allows this parameter to be measured continuously for each investor by monitoring transactions; therefore, correction can be made at any time. The optimal choice also depends on the possibilities investors may have; therefore, our model uses the calculated *EDR-E(r)* pairs, their distributions to adjust the leveraged efficiency frontier, the borrowing constraint and the interest rate for each investor. In our regression the iso-utility functions (using "a" CARA) and the efficiency frontier (using *x* leverage limit, $r_C$ interest rate and *EDR-E(r)* pairs), thus the optimal choice in their tangent point, can be defined for every investor. Figure 14 illustrates the choices where *A, B* and *C* are unleveraged, *A', B'* and *C'* are leveraged positions, O is the optimal choice and D is the interest rate (which is risk-free, thus *EDR=E(r)=$r_c$*). For different leverage possibilities or interest rates, investors' preference may change. In this case, Efficiency frontier 2 included both *A* and *B* portfolios but not *C*; however, for an investor with Efficiency frontier 1, *A'* and *C'* are efficient and *B'* is not.

### Figure 14: Individual optimization with leverage constraints

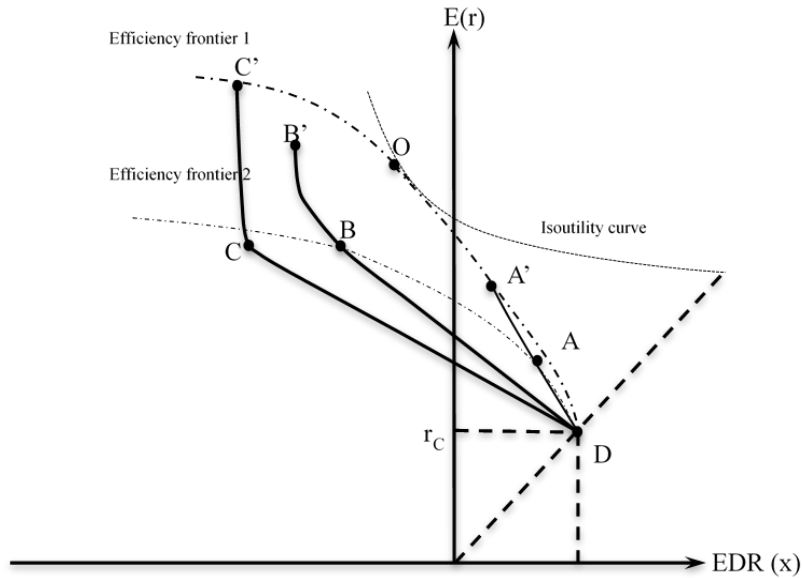

Notes: The graph represents three portfolios with different return distribtuions. As in Figure 13, the efficiency frontier may change as the leverage differs. Furthermore, combining the frontier with the highest reachable isoutility curve yields the optimal portfolio choice.

## 6. EXPECTED RETURN

### 6.1. Formation of the expected return

Until this point we assumed that investors are price-takers, thus one can invest in portfolios with constant, exogenous *EDR-E(r)* parameters. However, in the previous section we showed that available portfolios (and thus the efficiency frontier) can be fairly different for various investors, especially if we accept that some of them have the possibility of investing in highly leveraged positions. Therefore, hereafter we define individuals as price-maker participants of the market, although their effect on prices may differ significantly.

Institutions (such as brokerage firms) that have access to investors' trading data are able to estimate their behaviour based on historical actions. They can define their clients' risk aversion, their CARA and their utility function. In fact, every single parameter is known to make estimations on future *EDR-E(r)* pairs. These approximations are assigned to each investor to define their optimal choices. Furthermore, by aggregating these choices they can estimate the aggregate expected return of the market (which moves the prices). However, in order to have an aggregation, these institutions have to have a model to describe the relation between different required returns assigned to a given *EDR*. Distinct groups of investors may have different return requirements (especially due to leverage and interest rate differences); however, since every publicly traded asset has a unique price, these required returns have to add up to a unique expected return based on a function. Although the main purpose of this paper is not to define this precise function, in order to approximate a regression we use the value-weighted aggregating function. Thus, we add up each investor's required return with weights equal to their invested value to get a unique expected return.

This weighted aggregation function is based on macroeconomic demand and supply functions, where the price assigned to a portfolio is defined by the current aggregated supply and demand on the investment opportunity. Since databases enable analysis of who wants to make transactions at the current price with what volume and what the required return of the given investor is, the future price (current price multiplied by the required return) and the future volume of each investor can be defined. This way the aggregated supply (AS) function is known. According to Fama (1991), investors insist on smoothing their consumption over time; therefore, it seems to be realistic to assume that their incoming cash flows are balanced and continuous due to diversification. Furthermore, if the future cash flows are assumed to be fixed (which seems to be true according to the efficient market hypothesis by Fama (1970), in which prices reflect all the available information), the aggregated demand (AD) grows at the same pace as AS, which is the required return of investors. This means that in the discounted cash flow pricing method the exponent of the cost of alternative choice gets smaller over time. Combining the functions mentioned above we create the AS-AD system in our model, hence, one can approximate future prices of assets and their expected returns. An illustration of this approximation is shown in Figure 15, where the initial *(t₀)* AS and AD functions are increased by continuous return over time (represented on axis "*t*"). The functions used are $P_S=2Q$ and $P_D=5-2Q$, however, applying different ones does not change the results as long as aggregate supply and demand are increasing and decreasing functions of price respectively. We fix the continuous return at 20% in order to be expressive, hence, the exponential growth over time can clearly be seen.

**Figure 15: The formation of the expected return**

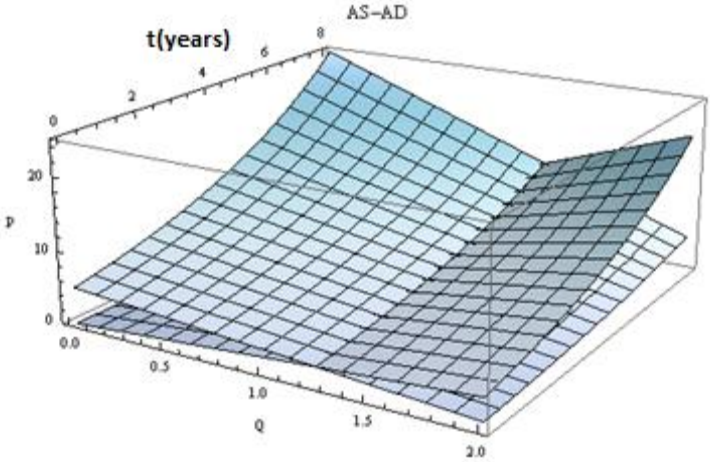

Notes: The simulated data represents the evolution of the price P as the function of aggregate demand and supply Q and time t. The increasing price shows the effect of the expected return over time.

We underline that this type of approximation of the expected return is very sensitive to the input data; therefore, the wider the sample is, the more precise the regression will be. The necessary data can be extracted by analyzing the actions of clients in financial institutions or brokerage services; however, using our model at national level (such as under the supervision of the Securities and Exchange Commission) or at international level (for example, with the administration of the International Monetary Fund (IMF) or the European Central Bank (ECB)) could produce fairly precise approximations of future prices of capital markets.

## 6.2. Empirical evidence for the EDR-E(r) relationship

Although, due to the lack of individual data, we cannot precisely define the aggregating function a brief empirical comparison with alternative risk measures can be made for a representative investor. Assuming that risk-free rate does not vary significantly and leverage limit is a function of risk expected return can be fitted on risk measures such as variance or EDR. The former assumption is necessary to define a unique constant in the regression, while the latter is required to be able to control for the selection of portfolios. An example for the latter would be a pattern that investors who are provided higher leverage limits choose portfolios with lower risk (hence lower variance or higher EDR) (see Appendix 3). Therefore, less risky assets are purchased by investors with higher leverage opportunities and higher risk-aversion, hence the price of assets is not determined by the correlation with a unique market portfolio but depends on the correlation with the reference portfolio. However, this reference is conditional on the leverage limit and risk-aversion, therefore, by assuming that capital markets include a diverse set of investors, no unique systematic risk (i.e. CAPM beta) exists as it is different for every single investor. Therefore, we argue that the risk premium should be described with the total risk and hence we apply the *EDR* measure.

In Table 2 we provide the empirical results of our illustration of *EDR*-based asset pricing. Here the overlapping yearly returns of the 340 S&P500 members traded both in 2009 and 2014 are applied in a linear estimation of the expeted return. The risk measures used are β, σ, semi-variance (σ-), downside beta (β_D) and *EDR*. The pricing equation of the EDR is estimated as

$$E(r) = 0.15 + 0.56 \cdot EDR. \qquad (30)$$

**Table 2: Linear estimations of the expected return**

| Risk measure | Coefficient | p-value | R-squared |
|---|---|---|---|
| Volatility | -0.1110 | 0.0466 | 0.01 |
| Semi-variance | -0.9198 | 0.0167 | 0.02 |
| Beta | -0.0181 | 0.0000 | 0.06 |
| Downside Beta | -0.0217 | 0.0006 | 0.03 |
| EDR | 0.5655 | 0.0000 | 0.62 |

Notes: Overlapping yearly returns on the 09/10/2009 to 09/10/2014 are applied. The coefficient indicates that of the applied risk-measure, p-value stands for its Student's-t test with the null of the coefficient being zero and the R-squared represents the coefficient of determination of the OLS estimations including a constant as well.

As this paper is a theoretical one, we consider these results promising based on the $R^2$ values and significance levels, however, due to limitation in length, we plan to discuss the detailed empirical test of the model in another paper.

## 7. CONCLUSION

Our proposed model yields novel results in four main fields of asset pricing. Firstly, we introduce a novel risk measure, Expected Downside Risk. Based on Conditional Value-at-Risk it keeps the advantages, such as subadditivity and convexity, but omits the disadvantages, such as subjectivity and the nongeneral approach to return distributions. This similarity is important since CVaR has gained high popularity in financial risk management nowadays (Bank for International Settlements (BIS), 2012), and therefore, switching to EDR could easily be done as they use very similar calculation methodology. Secondly, by combining Prospect Theory with Expected Utility Theory, our model is able to describe risk-seeking and risk-averse behaviour in one regression. Thirdly, we point out that the unlimited leverage assumption of existing equilibrium models can cause significant bias in asset price prediction. Using an alternative approach, our model may lead to more precise and unbiased estimations. And fourthly, through the aggregation function of individual required returns, the price-maker activity of investors can be modeled. Although, due to the lack of individual data, this function has not been tested in detail in this paper we provide a brief empirical estimation of the expected return with the use of aggregated data. There have been numerous attempts to describe this aggregating with market microstructural (Brennan and Subrahmanyam, 1996; Garman, 1976) or behavioural financial approaches (Shefrin and Statman, 1994), which, combined with our ideas, could pave the way for future research. Furthermore, we underline here that these theories mainly focus on the short-term changes of prices while, in the long term, our approximation of expected return at an aggregate level through EDR seems to work well in practice as individual effects are canceled out.

Moreover, we argue that equilibrium pricing should not be driven by factors with no fundamental background (although some of them may provide significant results for given periods) but should be based on well-defined theories. Our model proposes a method for asset pricing based on investors' preferences, hence, it explains the differences between expected returns with the diversity of groups investing in the given assets. Therefore, two distinct assets with equal volatility may provide different expected returns due to the difference in investors' opportunities and preferences. Although this means that expected return is not solely based on volatility, instead of including irrelevant factors in asset pricing models, future research on the topic should focus on which group of investors' trade the assets that have different characteristics and on what the reason is behind their choice.

Potential ways of future research could cover a detailed empirical analysis of the *EDR-E(r)* relationship and the comparison of its predicting power with alternative risk measures, where polynomial regressions could be tested as well. Another field of study would be the individual portfolio choice of investors either by analyzing the individual choice of the participants of capital markets or by testing whether the difference between the portfolio risk of investors accessing high and low leverage is significant (as suggested by Appendix 3). A third potential topic could be related to exmperimental analysis of investors' portfolio choice subsequent to gains or losses. Finally, as in-sample tests of *EDR* indicate extremely high goodness-of-fit relative to the alternatives, further in- and out-of-sample prediction models of the proposed risk measure could be tested as well using GARCH-type, VAR or other autoregressive models.

# Appendix 1

The VaR for a fixed α% shows the value that with *(1-α)%* probability the return of an investment will be above. An alternative explanation is that *VaR$_α$* is the *α* percentile of the density function of return. By defining *f(x,y)* as the function of loss where *x* is a chosen portfolio and *y* is a random variable, the probability of *f(x,y)* not exceeding a *ζ* value (loss) is *ψ*:

$$\psi(x,\zeta) = \int_{f(x,y)\leq\zeta} p(y)dy \qquad (A1.1)$$

Assuming a fixed *x* portfolio, this equation is the same as the cumulative distribution function of the *f(x,y)* loss function in *ζ*. According to this function, the Value-at-Risk (*VaR$_α$*) is:

$$\zeta_\alpha(x) = \min\{\zeta \in R: \psi(x,\zeta) \geq \alpha\} \qquad (A1.2)$$

meaning the first point on the cumulative distribution function of the return of the *x* portfolio with greater cumulative probability than *α*. *CVaR$_α$* can be described with the same technique:

$$\varphi_\alpha(x) = (1-\alpha)^{-1} \int_{f(x,y)\geq\zeta_\alpha(x)} f(x,y)p(y)dy \qquad (A1.3)$$

where $f(x,y) \geq \zeta_\alpha(x)$ is *(1-α)* according to the definition of *VaR*, hence it becomes the denominator. The interpretation of eq. (A1.3) is that *VaR$_α$(x)* is the *α* percentile of the

*f(x,y)* loss function, while $CVaR_α(x)$ is the probability-weighted average (expected value) of losses greater than *VaR*. Therefore, $CVaR_α(x) ≥ VaR_α(x)$ is always true (Rockafellar and Uryasev, 2002).

# Appendix 2

First, in order to precisely simulate risk aversion, we have to take into consideration that the modified utility curve may have other risk aversion coefficients by measuring the changes of wealth instead of the total wealth. Hence, we follow the method of Barsky et al. (1997) but instead equating

$$0.5U(xW) + 0.5U(2W) = U(W) \qquad (A2.1)$$

we measure

$$U(-(1-x)E(r)) = U(E(r)). \qquad (A2.2)$$

Plugging in the modified utility function we get

$$2.25U(1 - e^{-a(1-x)E(r)}) = 1 - e^{-aE(r)}, \qquad (A2.3)$$

and thus we get the risk aversion coefficient as

$$a = \ln \frac{1.25}{e^{-E(r)}(e^{1-x}-1)}. \qquad (A2.4)$$

Table A/1 represents the results of our Monte Carlo estimation for the risk aversion coefficients where the rows and columns stand for the expected return and the x values used by Barsky et al. respectively.

**Table A/1 Risk aversion coefficients of the modified utility function**

| x/E(r) | 50% | 66.70% | 80% | 90% | 92% | 95% |
|---|---|---|---|---|---|---|
| 1% | 0.665896 | 1.16164 | 1.740915 | 2.485312 | 2.718606 | 3.203772 |
| 2% | 0.675896 | 1.17164 | 1.750915 | 2.495312 | 2.728606 | 3.213772 |
| 3% | 0.685896 | 1.18164 | 1.760915 | 2.505312 | 2.738606 | 3.223772 |
| 4% | 0.695896 | 1.19164 | 1.770915 | 2.515312 | 2.748606 | 3.233772 |
| 5% | 0.705896 | 1.20164 | 1.780915 | 2.525312 | 2.758606 | 3.243772 |
| 6% | 0.715896 | 1.21164 | 1.790915 | 2.535312 | 2.768606 | 3.253772 |
| 7% | 0.725896 | 1.22164 | 1.800915 | 2.545312 | 2.778606 | 3.263772 |
| 8% | 0.735896 | 1.23164 | 1.810915 | 2.555312 | 2.788606 | 3.273772 |
| 9% | 0.745896 | 1.24164 | 1.820915 | 2.565312 | 2.798606 | 3.283772 |
| 10% | 0.755896 | 1.25164 | 1.830915 | 2.575312 | 2.808606 | 3.293772 |
| 11% | 0.765896 | 1.26164 | 1.840915 | 2.585312 | 2.818606 | 3.303772 |
| 12% | 0.775896 | 1.27164 | 1.850915 | 2.595312 | 2.828606 | 3.313772 |
| 13% | 0.785896 | 1.28164 | 1.860915 | 2.605312 | 2.838606 | 3.323772 |
| 14% | 0.795896 | 1.29164 | 1.870915 | 2.615312 | 2.848606 | 3.333772 |
| 15% | 0.805896 | 1.30164 | 1.880915 | 2.625312 | 2.858606 | 3.343772 |
| 16% | 0.815896 | 1.31164 | 1.890915 | 2.635312 | 2.868606 | 3.353772 |
| 17% | 0.825896 | 1.32164 | 1.900915 | 2.645312 | 2.878606 | 3.363772 |
| 18% | 0.835896 | 1.33164 | 1.910915 | 2.655312 | 2.888606 | 3.373772 |
| 19% | 0.845896 | 1.34164 | 1.920915 | 2.665312 | 2.898606 | 3.383772 |
| 20% | 0.855896 | 1.35164 | 1.930915 | 2.675312 | 2.908606 | 3.393772 |

It is clear that the results, although somewhat different, are not that far from those measured by Barsky et al. The risk aversion coefficients are thus in the same order of magnitude and can be used for describing portfolio choice.

Therefore, based on these coefficients, we are able to apply the $EDR(x) + \frac{0.8^2}{a} < E(r_x)$ constraint of eq. (23) to the simulation presented in Figure 5, where $a = 3$ is used.

**Figure A/1 Slope of the iso-utility curve**

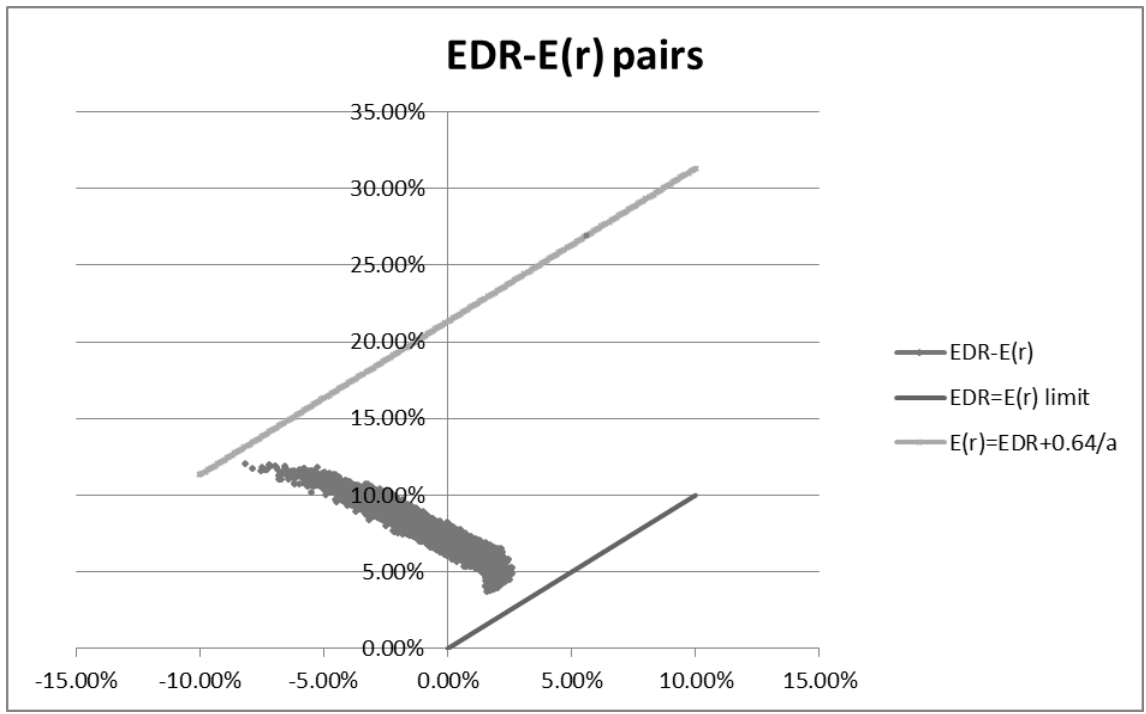

Even by considering this extremely risk-averse case of $a = 3$, portfolios can be found below the constraint, hence, marginal iso-utility curves at these points have negative slope and thus optimal choice can be found.

# Appendix 3

Without the loss of generality the assumption can be made that the efficiency frontier σ-E(r) pairs are described by a concave power function. Let us define this function as

$$E(r) = (\sigma - \beta)^\alpha + \gamma : \beta > 0,\ 0 < \alpha < 1. \qquad (A3.1)$$

The utility maximizing investor then has the problem of

$$\max_{\sigma, E(r)} U \ \ s.t.\ E(r) = (\sigma - \beta)^\alpha + \gamma, \qquad (A3.2)$$

which yields the solution of

$$\frac{dU}{d\sigma_{opt}} - \lambda \alpha (\sigma - \beta)^{\alpha-1} = 0, \qquad (A3.3)$$

$$\frac{dU}{dE(r)_{opt}} + \lambda = 0, \qquad (A3.4)$$

$$\sigma_{opt} = \frac{\alpha(\sigma_{opt} - \beta)^{\alpha-1}}{a}, \qquad (A3.5)$$

where $\sigma_{opt}$ stands for the portfolio risk at the optimum. We also know that the leveraged position can be defined as

$$E(r)_L = r_f + L\big(E(r) - r_f\big);\ \sigma_L = L\sigma, \qquad (A3.6)$$

where L stands for the leverage (e.g. L=1 with no leverage). The latter gives the solution of

$$\sigma_{L,opt} = \frac{\alpha(\sigma_{opt}-\beta)^{\alpha-1}}{La}. \qquad (A3.7)$$

The difference between the leveraged and unleveraged optmization is shown in Figure A/2, where equations (A3.5) and (A3.7) are represented by the cross-sections. The solid, dashed and dotted lines stand for the left hand side of equation (42) and the right hand sides of equations (A3.5) and (A3.7) respectively. The parameters are fixed at {α,β,a,L} = {0.5,0.1,4,2}.

**Figure A/2 Slope of the iso-utility curve**

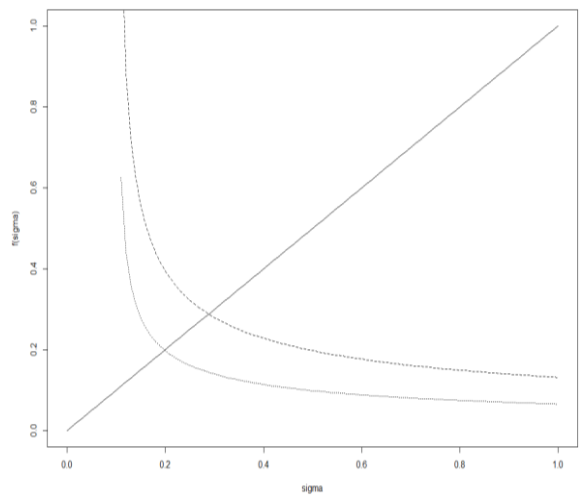

The analytical and graphical interpretations indicate that investors accessing higher leverage limit or being more risk-averse (i.e. *a* is high) choose portfolios with lower risk that is $\sigma_{L,opt} < \sigma_{opt}$.